\documentclass[]{aa}
\usepackage{graphicx}
\newcommand{\8}{3EG~J1828$+$0142}
\newcommand{\7}{3EG~J1735$-$1500}
\newcommand{\cc}{GRO~J1411$-$64}
\begin{document}
\title{A microquasar model applied to unidentified gamma-ray sources}
\author{V. Bosch-Ramon\inst{1} \and J.~M. Paredes\inst{1} 
\and G. E. Romero\inst{2,3,}\footnote{Member of CONICET} 
\and D. F. Torres\inst{4}}
\institute{Departament d'Astronomia i Meteorologia, Universitat de Barcelona, 
Av. Diagonal 647, E-08028 Barcelona, Spain; vbosch@am.ub.es, jmparedes@ub.edu.
\and Instituto Argentino de Radioastronom\'{\i}a, C.C.5,
(1894) Villa Elisa, Buenos Aires, Argentina; romero@iar.unlp.edu.ar.
\and Facultad de Ciencias Astron\'omicas y Geof\'{\i}sicas, UNLP, 
Paseo del Bosque, 1900 La Plata, Argentina.	
\and Lawrence Livermore National Laboratory, 7000 
East Avenue, L-413, Livermore, CA 94550; dtorres@igpp.ucllnl.org.}
\authorrunning{Bosch-Ramon et al.}
\titlerunning{A microquasar model for gamma-ray sources}
\offprints{V. Bosch-Ramon \\ }
\abstract{Among unidentified gamma-ray sources in the galactic plane, 
there are some that present significant variability and have been proposed 
to be high-mass microquasars. To deepen the study of
the possible association between variable low galactic latitude
gamma-ray sources and microquasars, we have applied a leptonic jet model based on the microquasar 
scenario that reproduces the gamma-ray spectrum of three unidentified gamma-ray sources, \7, \8\ 
and \cc, and is consistent with the observational constraints at lower energies. We conclude that 
if these sources were generated by microquasars,
the particle acceleration 
processes could not be as efficient as in other objects of this type 
that present harder gamma-ray spectra. Moreover, 
the dominant mechanism of high-energy emission 
should be synchrotron self-Compton (SSC) scattering, and the radio jets may only be observed at 
low frequencies. For each particular case, further predictions 
of jet physical conditions and variability generation mechanisms
have been made in the context of the model.
Although there might be other candidates able to explain the emission coming from these sources, microquasars 
cannot be excluded as counterparts. Observations performed by the next generation of gamma-ray instruments,
like GLAST, are required to test the proposed model.
} 

\maketitle
       
\section{Introduction} \label{intro}

The instruments EGRET\footnote{http//cossc.gsfc.gov/egret/} and COMPTEL\footnote{http//cossc.gsfc.gov/comptel/},
onboard the Compton Gamma Ray Observatory (CGRO),  detected about two hundred gamma-ray sources that still remain
unidentified.  Among these sources, there is a subgroup that appears to be concentrated towards the galactic
plane and presents significant variability (Torres et~al. \cite{Torres01}, Nolan et~al. \cite{Nolan03}). The
discovery of the microquasar LS~5039, a high-mass X-ray binary (XRB) with  relativistic jets, and its association
with the high-energy gamma-ray source 3EG~J1824$-$1514 (Paredes et~al. \cite{Paredes00}), opened the possibility
that some other unidentified EGRET sources (Hartman et~al. \cite{Hartman99}) could also be microquasars. 
That microquasars can be high-energy  gamma-ray emitters has been confirmed by the ground-based
Cherenkov telescope HESS, that detected a TeV source whose  very small 3-$\sigma$ error box contains LS~5039
(Aharonian et~al. \cite{Aharonian05}). In addition, high-mass microquasars have been proposed to be counterparts
of at least a significant fraction of the low galactic latitude unidentified
variable EGRET sources (e.g. Kaufman
Bernad\'o et~al. \cite{Kaufman02}, Romero et~al. \cite{Romero04a}). Recent statistical and theoretical studies on
this group of sources have provided additional support to this association (Bosch-Ramon et~al. \cite{Bosch05a}).
Therefore, it seems at least plausible  that microquasars could represent a significant fraction of the variable
gamma-ray sources in the galactic plane, generating not only the emission detected by EGRET but also that of
variable sources detected  by other gamma-ray instruments like COMPTEL. This paper deepens the study  of the
gamma-ray source/microquasar connection by applying a detailed microquasar model to three unidentified gamma-ray
sources: \7\ and \8, two likely variable unidentified EGRET sources in the galactic plane\footnote{
\7\ and \8, at galactic latitudes 9$^{\circ}$ and 6$^{\circ}$ respectively and assuming galactic distances, 
are at few hundreds of parsecs above
the galactic plane. It does not preclude that they are relatively young objects provided that the 
EGRET microquasar LS~5039 is a runaway object that could during its lifetime reach 
galactic latitudes up to 10$^{\circ}$ or 
vertical distances of 500~pc from the galactic plane (Rib\'o et~al. \cite{Ribo02}).} (Torres et~al. \cite{Torres01}, Nolan et~al. \cite{Nolan03}),  and \cc, recently discovered
by Zhang et~al. (\cite{Zhang02}) in a re-analysis of the COMPTEL data, which is also both variable and located in
the galactic plane. Our aim is to check whether a microquasar model "under reasonable assumptions" can be
compatible with the observational constraints at different frequencies.

The contents of this paper are arranged as follows: in Sect.~\ref{model}, the microquasar  model is
described; in Sect.~\ref{gamma}, the application of the model to each source as well as a brief
discussion of its results and predictions are presented; the work is summarized in
Sect.~\ref{con}. 

\section{The microquasar model} \label{model}

A semi-analytical model to calculate a microquasar spectrum from radio to gamma-rays has been developed
(Bosch-Ramon et~al. \cite{Bosch05a}). The scenario consists of an X-ray binary system where the compact
object, a black hole or a neutron star, surrounded by an accretion disk and a corona, generates collimated outflows or jets
(Mirabel \& Rodr\'iguez \cite{Mirabel99}). The photon fields originating in the companion star and the corona
(McClintock \& Remillard \cite{McClin04}) are  taken into account. The jet is modeled as an inhomogeneous and
magnetized relativistic flow of protons and leptons, and relativistic leptons dominate the radiative
processes. Protons will be important dynamically, and this has been taken into account in determining the
leptonic luminosity of the jet. This means that the total jet power cannot be less than 10 times the leptonic
power, since otherwise the conversion of the jet kinetic luminosity into radiation luminosity probably would be too
efficient (see Fender \cite{Fender01}). This fact, related to the macroscopic energy conservation law, imposes
that the accretion energy budget should be enough to power the whole jet (as seems to be the case in general,
see Bosch-Ramon et~al. \cite{Bosch05a}). Since it is not clear to what extent they are
relevant, we have not accounted for proton radiative properties. We refer to the work of Romero et~al.
(\cite{Romero03}) for the radiative properties of hadronic jets in microquasars.  

In this leptonic model, radio emission is generated by an outer jet that expands 
at a lower velocity than what is expected for the conical case. 
This type of expansion is introduced to simulate the particle re-acceleration
processes allowing extended radio
emission (i.e. through bulk motion dissipation of energy caused by external medium
interaction, or by instabilities in the flow of internal origin). 
This radio jet starts where the high energy jet emission 
is no longer significant, at about 100 times the distance of the jet injection 
point to the compact object.
Other works that have adopted slowly expanding jet models are, for instance, Ghisellini
et~al. (\cite{Ghisellini85}) and  Hjellming \& Johnston (\cite{Hjellming88}). In the optical-UV
band, emission is in general dominated by the star and, at higher energies, the
corona and/or the inner region of the jet. Because of the higher density and pressure conditions than those
of radio jets, this inner region is modeled as conical, i.e. undergoing free expansion. 

Jet particles interact with the present photon fields (synchrotron, star, accretion disk and corona
photons) through the inverse Compton (IC) effect.  In our case, the contribution from the disk scattered
photons is negligible in front of the corona IC component, since disk photons come from behind the jet and
few of the disk scattered photons reach the observer (Dermer et~al. \cite{Dermer92}).  Disk emission
itself cannot be particularly significant since it is constrained by the fact that the source remains
unidentified at X-rays (see below). Therefore, since this component is superfluous for the modeling, it has
not been considered.  We have accounted for both Thomson and Klein-Nishina regimes of IC interaction
(Blumenthal \& Gould \cite{Blumenthal70}). The different functions that represent the electron energyi
distribution, the maximum electron energy and the magnetic field within the jet have been parametrized 
to simulate their evolution along the jet (e.g., Ghisellini et~al. \cite{Ghisellini85}, Punsly et~al.
\cite{Punsly00}). 

\begin{figure}
\resizebox{\hsize}{!}{\includegraphics{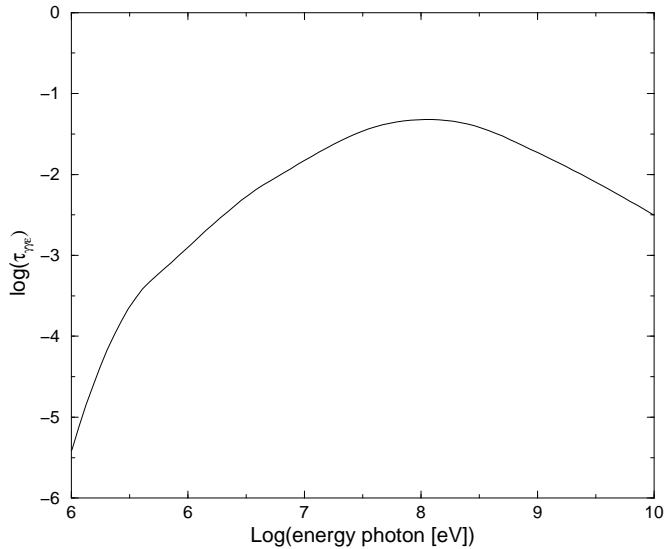}}
\caption{Opacities at different photon energies in the base of the jet. 
The dominant corona luminosity has been taken to be $3\times10^{34}$~erg~s$^{-1}$.}
\label{Fig:opac}
\end{figure}

For this model, the opacities due to pair creation under the jet conditions considered here have been computed
and are negligible at this stage. In Fig.~\ref{Fig:opac}, the opacities for different energies in the  base of
the jet, where they are the highest, are shown. The calculation has been performed for a general  case, similar
however to those treated below. Moreover, the opacity within the stellar photon field  at the photon energies
involved here is negligible. Nevertheless, for microquasars with spectra extending to 100~GeV, or with more
luminous corona and/or jet inner regions (e.g. see Romero et~al. \cite{Romero02}), opacities would not be
negligible.

Our model predicts variability through variations in the leptonic jet power, likely linked to
orbital eccentricity and accretion rate changes, as well as to changes in the jet viewing angle due to jet
precession. Changes in the jet kinetic luminosity imply an increase of the jet radiating  particle density, and
precession implies variation in the Doppler boosting that has implications for both the flux and the
maximum energy of the observed photons. This issue is discussed qualitatively in Sect.~\ref{impmqm}.

\section{Application of the model to unidentified $\gamma$-ray sources} \label{gamma}

In this section, we investigate whether a high-mass microquasar model could reproduce the gamma-ray emission
from \7, \8\ and \cc, observing other constraints at lower energies. We do not intend either to identify the
counterparts or to perform a statistical approach for fitting of our model because only higher resolution
gamma-ray observations can solve the identification problem and, regarding the latter issue, available data
are sparse and a statistical fit would be meaningless. 
We do not adopt 
the flux of any particular source in the gamma-ray error boxes 
as the reference one provided the counterpart is unknown, but
typical radio and X-ray fluxes of the sources found 
inside those fields are taken as upper limits at these energy ranges.   
If the emission at these frequency bands significantly overcame
the typical fluxes found in the gamma-ray error boxes, say, by one order 
of magnitude, the source would be barely unidentified. 
Lower-limits on the
fluxes cannot be stated since the counterpart could be relatively quiet in radio and X-rays, being unnoticed 
by the surveys carried out so far over the regions corresponding to the gamma-ray error boxes. All this implies that 
the flux can only be constrained roughly.

In the optical band, even though high-mass microquasars have bright stellar
companions, clear counterparts have not been found in the gamma-ray error boxes. This could be explained
by the strong absorption and/or enshrouding in the optical, UV and even in the X-ray band that is often
present towards the galactic plane. For instance, it has been suggested that obscured
INTEGRAL\footnote{http://integral.esac.esa.int/} sources could be intrinsically
or locally obscured in the UV and X-ray band (e.g.
Walter et~al. \cite{Walter03}). Furthermore, emission from the massive companion of an X-ray binary 
scattered and/or reprocessed to the far infrared could even be too weak to be detected by, for instance, the
satellite IRAS (e.g. Filliatre \& Chaty \cite{filli04}). At the adopted distances, the bright
companions assumed here would present a relative brightness in the optical band of about 12 magnitudes, if not
absorbed.

In the absence of specific knowledge, we have fixed the values of the jet parameters entering in the model to
fiducial standards for microquasars. For the binary system parameters and jet size, we have adopted
those used in Bosch-Ramon et~al. (\cite{Bosch05a}), and a Lorentz factor of 1.2, similar to that shown by 
the microquasars LS~5039 (3EG~J1824$-$1514, Paredes et~al. \cite{Paredes00}) and LS~I~+61~303
(3EG~J0241$+$6103, Kniffen et~al. \cite{Kniffen97}), which present mildly relativistic jets (Paredes
et~al. \cite{Paredes00}, Massi et~al. \cite{Massi04}). This should be sufficient, if
gamma-ray microquasars in the low-hard state share similar properties. For the jet viewing angle
($\theta$), provided again that the jets of LS~5039 and LS~I~+61~303 are mildly relativistic and it is not
required for them to have small $\theta$ to be detected (Paredes et~al. \cite{Paredes00}, Massi et al.
\cite{Massi04}), we have taken a mean value in our specific models of $45^{\circ}$.  
For the corona spectrum, we have adopted a power-law plus an exponential cut-off with the
maximum flux at 100~keV. The star has been taken to be a black-body peaking at UV energies.

The electron energy distribution (assumed to be a power-law of index $p$ and starting from energies of
about 1~MeV) and the corona luminosity have been chosen such that they reproduce the gamma-ray
spectra and are compatible with the fluxes at lower energies, adopted similar to those inferred from
typical sources in  the error boxes. As stated above, disk emission itself is limited by the X-ray
observational constraints which, together with a lower IC scattering probability (Dermer et~al.
\cite{Dermer92}), makes its IC contribution negligible (for the assumed viewing angle of 45$^{\circ}$, it
is almost one order of magnitude smaller than the corona IC contribution). The electron maximum energy
together with the magnetic field,  given a certain value of $p$, have been taken to reproduce properly
the observed spectral slope of gamma-rays. Since the spectral EC components seem to be unable to
reproduce the spectrum in gamma-rays, the magnetic field has been taken such that the SSC process is dominant.
For instance, if the corona scattered photons were dominant, it would violate the X-ray constraints for any
reasonable parameter choice. The leptonic jet power has been taken to obtain the luminosities
expected if the sources are at one particular distance (see below). However,
the adopted value is similar to those obtained for microquasar jets in previous works using different
approaches (e.g. Bosch-Ramon \& Paredes \cite{Bosch04a}).  Although specific values are provided in
Table~\ref{common}, we give in Sect.~\ref{impmqm} the set of values for the magnetic field, the jet power
and the electron maximum energy that are compatible with data.

The distance from these sources to the Earth has been taken to be $\sim$4~kpc. We have 
assumed that the sources are located close to the inner spiral arms, which have been
associated with microquasar birth regions (Bosch-Ramon et~al. \cite{Bosch05a}). 
To investigate the variability
properties of the studied sources within the context of our model, we have computed the spectral energy
distributions (SEDs) associated with the average and the maximum level of the observed gamma-ray 
fluxes. In the case of the two EGRET sources, the average
flux (luminosity) is given by the total exposure EGRET spectrum
\footnote{http://cossc.gsfc.nasa.gov/compton/data/egret/}, 
and the maximum flux (luminosity) is given by the highest flux
among the different EGRET viewing period fluxes (Hartman et~al. \cite{Hartman99}). 
To extrapolate fluxes at lower energies, we have assumed that the variations are linked to changes in
the accretion rate, linearly related to the jet power, although it is possible to distinguish jet power 
variations from precession (see Sect.~\ref{impmqm}).
For the COMPTEL
source, the average value and the maximum one are very similar because actual detections 
are similar in flux, and the remaining observations only were able to give upper limits for the source
(see Zhang et~al. \cite{Zhang02}). 

\subsection{3EG J1735$-$1500}

\begin{figure}
\resizebox{\hsize}{!}{\includegraphics{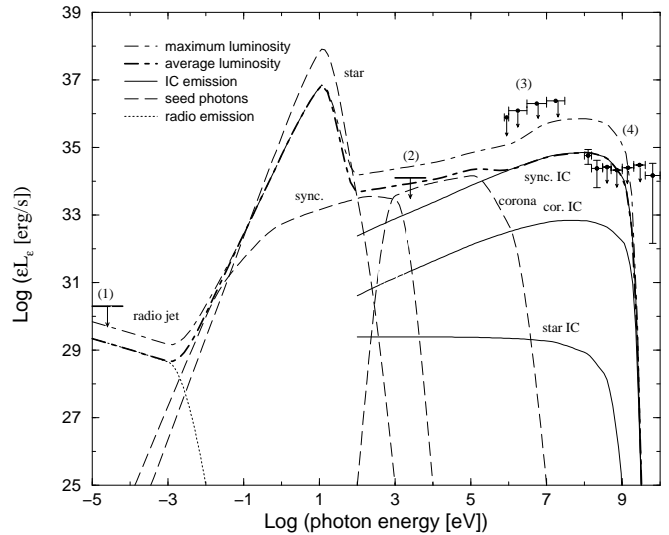}}
\caption{SED for a broadband microquasar model of the source 3EG J1735$-$1500. 
The SED of the average as well as the maximum luminosity level of the source are shown.
The adopted values for the different parameters are shown in Table \ref{common}.  
Upper limits at radio (1), X-ray (2) and COMPTEL (3) energies, as well 
as the average EGRET spectrum (4), are presented. To compute the total emission, 
the star component has been reduced a certain factor, in accordance to the maximum visual extinction 
found in the direction of the EGRET source. For the UV, we have followed roughly the relationship between 
different wavelengths provided by Valencic et~al. (\cite{Valencic04}).}
\label{Fig:1735}
\end{figure}

\begin{table*}[]
  \caption[]{Common and specific values for the parameters}
  \label{common}
  \begin{center}
  \begin{tabular}{cl}
  \hline\noalign{\smallskip}
Parameter &  values \\
  \hline\noalign{\smallskip}
Stellar bolometric luminosity [erg~s$^{-1}$] & $10^{38}$ \\
Distance from the apex of the jet to the compact object [cm] & $5\times10^7$ \\  
Initial jet radius [cm] & $5\times10^6$ \\
Orbital radius [cm] & $3\times10^{12}$  \\ 
Viewing angle to the axis of the jet [$^{\circ}$] & $45$ \\ 
Jet Lorentz factor & 1.2 \\
\end{tabular}
\begin{tabular}{llll}
\hline\noalign{\smallskip}
 & 3EG~J1735-1500&3EG~J1828+0142&\cc \\
\hline\noalign{\smallskip}
Jet leptonic kinetic luminosity [erg~s$^{-1}$] & $5\times10^{34}$&$10^{35}$&$3\times10^{35}$ \\
Maximum electron Lorentz factor (jet frame) & 3$\times10^3$&4$\times10^3$&5$\times10^2$ \\
Maximum magnetic field [G] & 10000&5000&8000 \\
Electron power-law index & 2&2&1.5 \\  
Total corona luminosity [erg~s$^{-1}$] & 3$\times10^{34}$&$3\times10^{33}$&$3\times10^{33}$ \\
  \noalign{\smallskip}\hline
  \end{tabular}
  \end{center}
\end{table*}

\7\ was considered in the work of Torres et~al. (\cite{Torres01}) as a variable EGRET source, and in Nolan
et al. (\cite{Nolan03}) it was also among the group of likely variable EGRET sources (probability $\sim60$\%).  The EGRET spectrum 
shows a photon index
$\Gamma\sim3.2\pm0.5$ and  average flux $\sim3\times10^{-11}$~erg~s$^{-1}$~cm$^{-2}$. The error box of \7\ was
explored by Combi et~al. (\cite{Combi03}), who proposed two potential counterparts: a radio galaxy
(J1737$-$15) and a compact
radio source (PMN~J1738$-$1502), a blazar candidate, that presents a flat radio spectrum and flux densities of about 0.3~Jy.
However, since at the present stage it is still hard to explain both whether a radio galaxy can present the 
variability of \7\ and the absence of X-ray counterpart for the compact radio source, we have not considered
them as definitive solutions of the identification problem. To model the  SED of a microquasar that could be the
origin of the EGRET emission, we take into account  the known observational data and constraints  at different
wavelengths. If the distance were 4~kpc, the typical luminosities of the radio sources in the EGRET error box
would be of about 2$\times10^{30}$~erg~s$^{-1}$, the X-ray luminosities would be $\sim10^{34}$~erg~s$^{-1}$, and
at COMPTEL energies the upper limits  would be $\sim10^{36}$~erg~s$^{-1}$ (Zhang et~al. \cite{Zhang04}). The
used parameter values are presented in Table~\ref{common}. The computed SED for both the average and the maximum
luminosity levels of the gamma-ray source are shown in Fig~\ref{Fig:1735}. It appears that \7, even if
detectable at X-rays during its maximum luminosity level, would be faint at radio wavelengths. 
At optical wavelengths, we have computed the visual extinction of 1.4 magnitudes 
from the relationship with the hydrogen column
density found by Predehl \& Schmitt (\cite{Predehl95}). It seems from Fig.~\ref{Fig:1735} that  additional
intrinsic absorption would be necessary to obscure the source in the optical band to prevent an easy
identification, since it still has an absolute brightness of 13.4 magnitudes.
To reproduce the observed gamma-ray variability through the jet precession, with the adopted
mildly relativistic velocity of the jet, the variation in the angle should be large, reaching almost
$0^{\circ}$. However, an orbital eccentricity of 0.5 or less could be enough to change the jet power,
producing the observed ratio of maximum to average luminosity (see, e.g., Bosch-Ramon et~al. \cite{Bosch05b}).

\begin{figure}
\resizebox{\hsize}{!}{\includegraphics{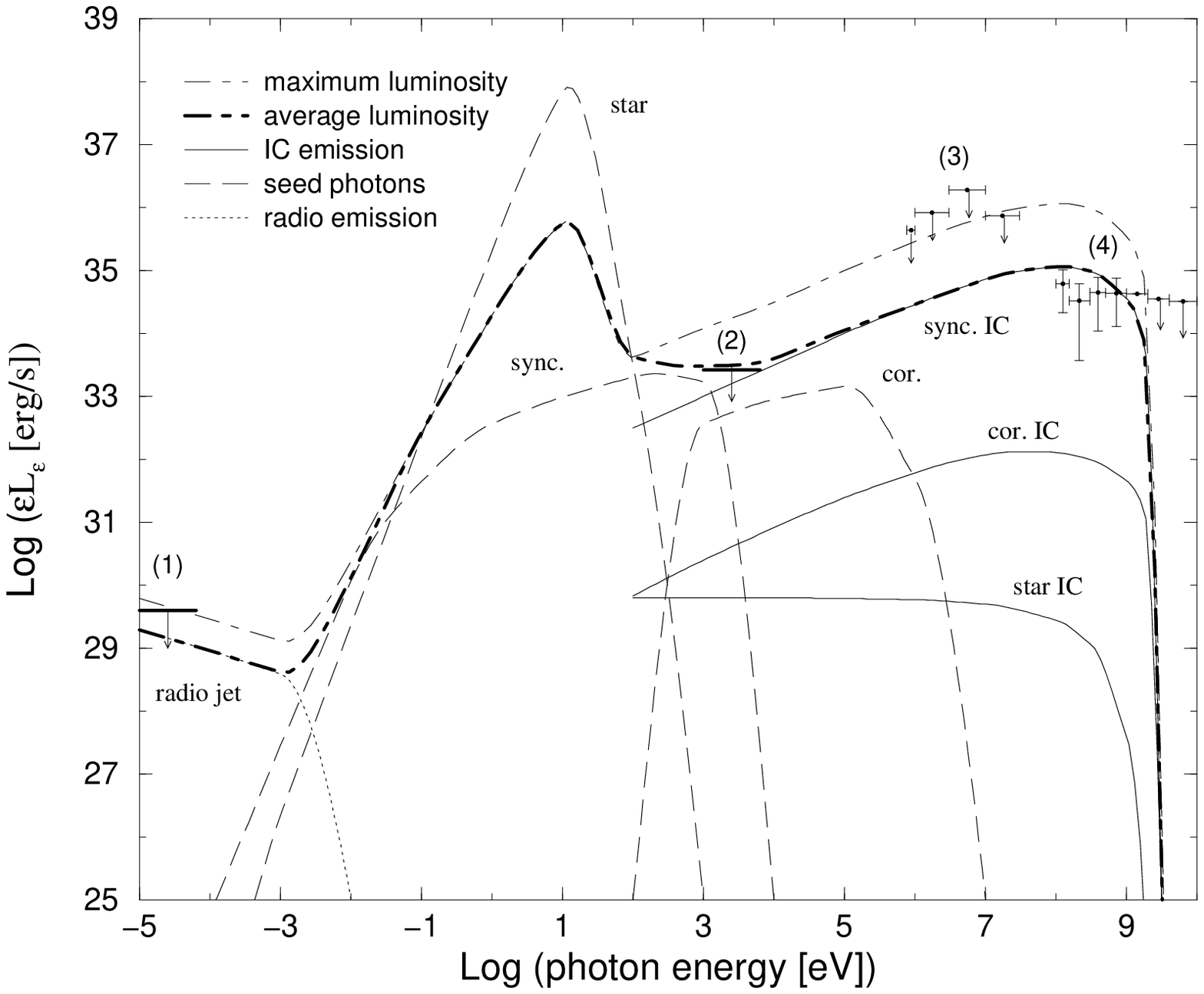}}
\caption{Same as in Fig.~\ref{Fig:1735} but for \8.}
\label{Fig:1828}
\end{figure}

\subsection{3EG J1828$+$0142}

\8\ is the second most variable low galactic latitude non-transient gamma-ray source in the list of 
variable EGRET sources of Torres et~al. (\cite{Torres01}), considered also very variable by 
Nolan et~al. (\cite{Nolan03}). The EGRET photon
index  is $\Gamma\sim$~2.7$\pm$0.4, with an average flux $\sim4\times10^{-11}$~erg~s$^{-1}$~cm$^{-2}$. Within
the error box of this EGRET source, there are several faint non-thermal radio sources 
with luminosities around 5$\times10^{30}$~erg~s$^{-1}$  (Punsly et al.
\cite{Punsly00}), and X-ray sources (observed by
the ROSAT\footnote{http://heasarc.gsfc.nasa.gov/docs/rosat/rosgof.html} All Sky Survey) with typical luminosities of about 10$^{33}$~erg~s$^{-1}$. COMPTEL upper
limits are also known (Zhang et~al. \cite{Zhang04}), corresponding to luminosities of about
10$^{36}$~erg~s$^{-1}$; the assumed distance still being the same. A supernova remnant (SNR), located at
$\sim$1~kpc, has been proposed by Punsly et al. (\cite{Punsly00}) to be associated with the
object emitting at gamma-rays. This SNR, yet not part of the Green's Catalog, was
not a member of the sample in the systematic study of molecular
material by Torres et al. (\cite{Torres03}), although the source
variability argues against a physical association with the SNR
shock.
Association with the SNR would imply a lower energy requirement
to explain the observed EGRET flux, although with such a distance
the source would not be associated with the Carina arm, 
as most of the EGRET sources in the galactic plane seem to be (Bhattacharya et~al. \cite{Bhattacharya03}). 
Also, there is a strong
flat spectrum radio source within the error box of this source which has been proposed to be a blazar
(Halpern et~al. \cite{Halpern03}; Sowards-Emmerd et~al. \cite{Sowards03} assigned to the blazar 
J1826$+$0149 an association probability of 60--80\%). 
Further observational data is needed for a firm association of the source with any
particular counterpart.
The values used for the different parameters are presented in Table~\ref{common} and the computed SED for both
the average and the maximum luminosity level of the gamma-ray source are shown in Fig~\ref{Fig:1828}.
It appears that the X-ray emission of \8\ could be at the detected source fluxes, and it
might be one of the radio sources in the EGRET error box during its most active state.
For an absorption of 2.6 magnitudes in the optical band, the optical counterpart would be
of a magnitude of about 15, which makes this source largely irrelevant from the optical point of view among 
other sources in 1$^{\circ}$-field. This will be 
more so in the ultraviolet, preventing a clear identification.
Regarding the variability, the same remarks made for the previous source are applicable to this case,
although the ratio of the maximum to average luminosity is slightly smaller and lower eccentricity and/or
precession could explain this finding.

\subsection{\cc}

\begin{figure}
\resizebox{\hsize}{!}{\includegraphics{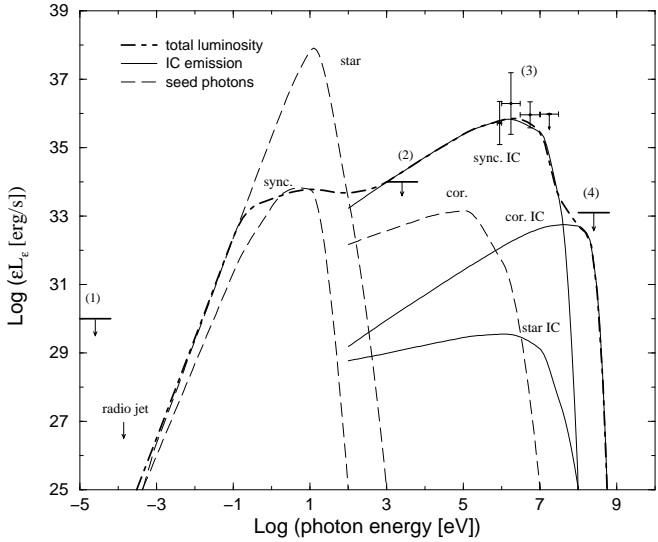}}
\caption{SED for a broadband microquasar model of \cc. 
The total COMPTEL spectrum (3), the same for the average and the maximum level of emission, 
as well as the 
upper limits at radio (1), X-ray (2) and EGRET energies (4).
The adopted values for the different parameters are shown in Table \ref{common}. The total emission 
has been reduced in the optical and ultraviolet bands according to the visual extinction in the COMPTEL 
source direction.}
\label{Fig:unCs}
\end{figure}

The detection by COMPTEL of a variable unidentified gamma-ray source in the galactic plane, \cc, was
reported by Zhang et al. (\cite{Zhang02}). This source has a photon index  $\Gamma\sim2.5\pm0.2$
and a flux of about $5\times10^{-10}$~erg~s$^{-1}$~cm$^{-2}$ at 10~MeV. The error box of COMPTEL is
large, with a radius of about 2 degrees. Several models have been proposed by Zhang et~al.
(\cite{Zhang02}) to explain the gamma-ray emission from this source
as a Be/X-ray binary (Romero et~al.
\cite{Romero01}), a weak galactic microblazar (Kaufman Bernad\'o et~al. \cite{Kaufman02}) or an
isolated black hole (Punsly et~al. \cite{Punsly00}). Inside the COMPTEL error box,
there are only two identified X-ray sources: 2S~1417$-$624, 
a transient Be/XRB pulsar (see Romero et~al.
\cite{Romero04b}), and GS~1354$-$645, a transient black-hole low-mass XRB. 
The remaining detected
X-ray sources have no counterparts at other wavelengths. These two XRBs lie just
inside the 4$\sigma$ region, and a physical association does not seem likely if the center 
of gravity of the source location is correct. 
The Circinus Galaxy is inside the error box though, if there are not ultraluminous gamma-ray  
objects in this galaxy, it is unlikely to be the counterpart of \cc. 
In the radio band, the typical flux of the sources found by
the PMN survey (Parkes-MIT-NRAO\footnote{http://www.parkes.atnf.csiro.au/research/surveys}) 
is taken as the upper limit at these wavelengths: a few 10~mJy, or about
$10^{30}$~erg~s$^{-1}$. For the constraints on the X-ray flux we will take a luminosity similar to
most of the sources detected by ROSAT, i.e. about $10^{34}$~erg~s$^{-1}$. The distance was
assumed to be 4~kpc. At high-energy gamma-rays, we will consider the sensitivity limit of EGRET in
the region of \cc\ as the upper limit. 
For this case, the average flux
and the maximum flux observed by COMPTEL are very similar (Zhang et al. \cite{Zhang02}).
The values used to compute the SED for the different parameters are presented in
Table~\ref{common} and the SED is shown in Fig~\ref{Fig:unCs}. 
As can be seen, the counterpart might be one of the X-ray sources detected in the
COMPTEL error box but its radio emission is too faint for detection. The visual extinction 
within the COMPTEL error box can reach 7 magnitudes. This could imply that  
intrinsic absorption is not required to preclude 
the detection of the optical and ultraviolet counterpart.

\subsection{Implications of the microquasar model} \label{impmqm}

\subsubsection{Source properties}
 
Our general conclusions are that, to reproduce the observed soft spectra at gamma-rays, a leptonic radiative
process  and a low maximum energy for the particles seem to be required. Generally, if the mechanism of emission were
hadronic, the  spectra would be harder. Moreover, comparing with two microquasar candidates likely
to be gamma-ray sources, LS~5039 and LS~I~+61~303, the electron maximum energies for these two cases 
(Bosch-Ramon \& Paredes \cite{Bosch04a}, \cite{Bosch04b}, Aharonian et~al. \cite{Aharonian05}) 
seem to be significantly higher than for the sources
treated here, likely pointing to a more efficient acceleration mechanism. In addition, if the sources were
microquasars, the dominant emitting process at high energies likely would be SSC. The dominance of SSC
scattering implies that the magnetic field is strong enough to obtain gamma-ray fluxes in agreement with
observations and preventing to increase the leptonic jet power to untenable values. This would be the case if
the magnetic field were too low and/or the corona scattered photons dominant. Within the context of the model,
the values for the magnetic field, the jet power and the maximum electron energy distribution can be
restricted to 10000~G, 10$^{35}$~erg~s$^{-1}$ and 1 GeV respectively.
Concretely, the COMPTEL source would present slightly higher 
jet power and lower maximum electron energy than the two EGRET sources. Otherwise the observed
spectrum at gamma-rays could not be reproduced taking into account the observational constraints and the
previous theoretical considerations. 
It is worth noting that these values are 
coarse estimates of the source properties under a microquasar assumption,
not being possible to achieve a better precision because of the lack of knowledge of the
counterpart fluxes at low energies.
Below 100 keV, the spectra must be hard enough to
agree with observations. This means that, while for \7\ and \8\ an electron power-law index of 2 is hard
enough, an index of 1.5 is required for \cc\ to keep the X-ray fluxes to those presented by the sources in the gamma-ray
error box. This could be related to a more relativistic shock acceleration in the particle injection of \cc,
and the lower maximum energy could be associated with stronger losses. We also note that the magnetic
field values are 100 times smaller than those of equipartition with relativistic electrons\footnote{Usually, it is considered 
to be around  equipartition in the
inner disk regions. However, this magnetic field is not known at the base of the jet (at $\sim10^8$~cm
from the compact object). Here it has been treated as a free parameter.}, which is about 10$^6$~G for a leptonic jet
power of
10$^{35}$~erg~s$^{-1}$. Finally, as noted above, due to
the stringent constraints in X-rays, the corona should be faint, which is in agreement with the moderate X-ray
emission as well as the lack of clear disk and corona features in the X-ray data of the two likely EGRET
microquasars LS~5039 (Bosch-Ramon et~al. \cite{Bosch05b}) and  LS~I~+61~303 (Leahy et~al. \cite{Leahy97}).

The radio jets associated with \7\ and \8 could only be detected if the electron energy losses are very low
and/or there is re-acceleration, perhaps due to shocks with the ISM at large scales or to internal shocks
caused by  different flow velocities (Marscher \& Gear \cite{Marscher85}). In such a case, in the context
of our model, there would be emission at low frequencies (below 1~GHz) up to large distances (about 1~pc).
To detect it would require an instrument with low angular resolution (about 1~arcmin) and high sensitivity
(about 0.1~mJy). For \cc, it seems that radio emission would not be detectable due to the low maximum 
electron energies and the strong losses in the inner jet. Therefore, these
microquasars, in contrast to what is usually expected, would not present clear radio jets. Instead, they
would present at most diffuse and faint radio lobes. 

\subsubsection{Variability}

The two mechanisms of variability that we have studied are leptonic jet power changes, associated with
accretion rate changes (e.g. for LS~5039, see Bosch-Ramon et~al.
\cite{Bosch05b}), and precession (e.g.
for LS~I~+61~303, see Massi et~al. \cite{Massi04}; for a general  case, see Kaufman Bernad\'o et~al.
\cite{Kaufman02}). We note that the plotted maximum luminosity SEDs for \7\ and \8\ below gamma-rays
correspond to those produced by the variation in the leptonic jet power. However, precession cannot be
discounted. \7\ and \8\ present average luminosities at  gamma-rays that are close to those of their
minima (Hartman et~al. \cite{Hartman99}), which could mean that the peaks are
short duration events (e.g.
periastron passage of an eccentric orbit or a minimum $\theta$ during the precession of the jet) on the
timescales of the EGRET viewing periods (of about two weeks). Instead, \cc\ shows a long duration burst
(Zhang et~al. \cite{Zhang02}), that could be more associated with a super accretion rate phase than to a
persistent jet affected by regular changes of its characteristics. The fact that this source appears to be
the brightest, assuming the same distance as for the rest, would give weight to this option. 

\subsubsection{Predictions}

In the radio band, a low resolution and high sensitivity instrument would be required to detect \7\ and
\8, expecting a very soft spectrum, whereas \cc\ would not be detectable. If this source is strongly
absorbed in the optical and the UV band, it could be still detectable in the infrared band, with higher
emission at longer than at shorter wavelengths. However, to test such statement, the location accuracy of
the sources should be improved, due to the large number of infrared sources within the gamma-ray error
boxes. At X-rays, \8\ could be detected with reasonable exposure times  (e.g. with
XMM\footnote{http://xmm.vilspa.esa.es/}), whereas \7\ and \cc\ would be easily detected due to their higher
emission levels at this energy band. For the three sources, the X-ray spectra would present photon indices
of 1.5 or less.  We note that XMM and INTEGRAL observations of \cc\ are underway, and we
will report on them elsewhere. Observations with the next generation gamma-ray instruments are
fundamental to properly associate the gamma-ray sources with their counterparts at lower energies. In the
COMPTEL energy range, \7\ and \8\ might be detected, at the adopted distance of 4~kpc, with an instrument
1--2 orders of magnitude more sensitive than COMPTEL. In the EGRET energy range, \cc\ might be detected by 
GLAST\footnote{http://glast.gsfc.nasa.gov/ssc/} with reasonable exposure times, if observed during an 
activity period similar to that presented during COMPTEL observations. Due to the very steep
cut-offs at energies beyond 1~GeV for \7\ and \8, and beyond 100~MeV for \cc, these sources would not be
detected by the new Cherenkov telescopes, although it does not prevent the detection of other microquasars
with higher maximum electron energies and/or more beaming. Finally, there are different observational
features depending on the dominant variability mechanism. A precessing jet would likely show a periodic
variation of both the photon index and the maximum detectable energy at gamma-rays. Moreover, the corona would
not suffer variations and gamma-rays (SSC) and radio (synchrotron) emission would vary in the same manner
(Dermer et~al. \cite{Dermer95}). If accretion is the origin of variability, SSC emission mechanism will
imply a different response to accretion changes than that presented at X-rays when dominated by the corona. 
In both cases, however, if our microquasar hypothesis is true, \7\ and \8\
variability should be periodic.

\section{Summary} \label{con}

A microquasar model is applied to model the emission at different wavelengths coming from the direction of
\7, \8 and \cc. In the context of this model, the gamma-ray emitting jets would radiate mainly via
SSC, and would present a lower electron maximum energy than the microquasars LS~5039 and LS~I~+61~303. 
Due to the low electron
maximum energy, the radio emission is low, and only detectable for the two EGRET sources at low
frequencies if the electrons are effectively re-accelerated in the radio jet, which is expected to be
quite extended due to the low radiative efficiency for the electron energy and magnetic field values 
in there. For the COMPTEL source, detectable radio emission is not expected. We have
estimated under what conditions the variability could be produced in the context of both precession and/or eccentric
orbit effects, although a scenario where both effects are present seems likely.

\begin{acknowledgements}

We thank Dr. Isabelle Grenier, referee of this paper, for useful comments and suggestions that 
have significantly improved this work.
We are grateful to Marc Rib\'o for his useful comments.
V.B-R. and J.M.P. acknowledge partial support by DGI of the Ministerio de Educaci\'on y Ciencia 
(Spain) under grant AYA-2004-07171-C02-01, as well as additional support from the
European Regional Development Fund (ERDF/FEDER). During this work, V.B-R has been supported
by the DGI of the Ministerio de (Spain) under the fellowship
BES-2002-2699. G. E. R is supported by Fundaci\'on Antorchas and the Argentine Agencies
CONICET and ANPCyT (PICT03-13291). 
The work of DFT was performed under the auspices of the US DOE (NNSA),
by UC's LLNL under contract No. W-7405-Eng-48.

\end{acknowledgements}

{}

\end{document}